# Optimized frequency comb spectrum of parametrically modulated bottle microresonators

M. Crespo-Ballesteros[1,3], A. B. Matsko[2], and M. Sumetsky[1]

[1]*Aston Institute of Photonic Technologies, Aston University, Birmingham B4 7ET, UK*
[2]*Jet Propulsion Laboratory, California Institute of Technology, Pasadena, CA, USA*
[3]*m.crespo@aston.ac.uk*

The formation of optical frequency combs (OFCs) by the parametric modulation of optical microresonators is commonly described by lumped-parameter models. However, these models do not consider the actual spatial distribution of the parametric modulation (SDPM). Here, we show that the effect of the SDPM becomes of special importance for an elongated SNAP bottle microresonator (SBM) having shallow nanometre-scale effective radius variation along its axial length. The advantage of SBMs compared to microresonators with different shapes (e.g., spherical and toroidal) is that SBMs, remaining miniature, can have resonant spectrum with much smaller free spectral range and no dispersion. Therefore, SBMs can be used to generate OFCs with much lower repetition rates. We consider the resonant and adiabatic modulation of parabolic SBMs and show that it is possible to improve the flatness and increase the bandwidth of the generated OFC spectra by optimising the SDPM. We suggest that the determined optimal SDPM can be experimentally realized using piezoelectric, radiation pressure, and electro-optical excitation of an SBM.

**Introduction**

The generation of optical frequency combs (OFCs) represents a vibrant field of research in photonics[1,2] with numerous applications in precision spectroscopy, optical metrology, and high-speed optical communications [3–6]. Currently, the main techniques to generate OFCs use mode-locked lasers[7], electro-optic (EO) modulation of continuous wave lasers[8], and nonlinear optical processes in microresonators[9,10]. The most common method to generate OFCs in microresonators is based on the nonlinear Kerr effect[11–14], but alternative approaches have been proposed, such as the application of the EO effect in microresonators made of materials with second-order nonlinearity[15–20]. The microresonator-based comb technologies are particularly attractive because they offer compact, efficient, and high-repetition-rate frequency combs (of the order of GHz and THz). However, applications such as high-precision spectroscopy require OFCs with lower repetition rates (a few hundreds of MHz or smaller)[3,21], which is more challenging to achieve with microresonators. Indeed, the repetition rate frequency of microresonator-based OFCs is usually equal to the microresonator free spectral range (FSR), $\Delta\nu_{FSR}$, which, for the commonly used ring, toroidal, and spherical microresonators, is inversely proportional to their size. As an example, a silica toroidal or spherical microresonator with $\Delta\nu_{FSR} = 50$ MHz have a radius $r_0 = c/(2\pi n_0 \Delta\nu_{FSR}) = 65$ cm (here $c$ is the speed of light and $n_0 = 1.46$ is the refractive index of silica). Therefore, to achieve sufficiently low repetition rates, the microresonator size has to be increased to dimensions which may be unpractical.

One approach to solve the challenge of generating OFCs with low repetition rates in microresonators consists in the adiabatic modulation of microresonator parameters at frequency $\nu_{par}$ that is much smaller than its FSR, $\nu_{par} \ll \Delta\nu_{FSR}$. In this case, the repetition rate of the OFC is equal to the modulation frequency and independent of the microresonator size. In a recent paper[22], the authors demonstrated a flat frequency comb spectrum with repetition rate as low as 50 MHz generated by optomechanical oscillations of a toroidal silica microresonator. The adiabatically slow modulation of the microresonator eigenfrequency depends on time as $\nu_e(t) = \nu_{e0} + \delta\nu_{par}\cos(2\pi\nu_{par}t)$, where $\delta\nu_{par}$ is the amplitude of modulation. This modulation can generate a close to uniform OFC formed by $N \cong \delta\nu_{par}/\nu_{par}$ spectral resonances separated by the modulation frequency $\nu_{par}$ (see consideration below). For a microresonator with radius $r_0$, the parametric modulation corresponds to an amplitude change of the effective radius variation (ERV) defined by the scaling relation $\delta r_{eff} = r_0\delta\nu_{par}/\nu_{e0}$. For the experimental parameters of ref.[22], $r_0 \cong 30$ μm, $\nu_{e0} \cong 200$ THz, $\nu_{par} = 50$ MHz, and $N \cong 1000$,, we find $\delta r_{eff} \cong 8$ nm. In ref.[22], this modulation was generated by the radiation pressure of the resonant high-power input light that excited the mechanical breathing mode of the microresonator.

The application of SNAP bottle microresonators (SBMs) with shallow nanoscale ERV, which are introduced at the surface of optical fibres[23–27], constitutes an alternative and promising approach to generate OFCs with low repetition rates. Optical eigenmodes in SNAP microresonators are whispering gallery modes (WGMs) adjacent to the fibre surface and slowly bouncing between turning points along the microresonator axial length. The adiabatic generation of OFCs with low repetition rates can be realized in SBMs by a strong input light, similar to the approach of ref.[22]. Alternatively, the EO modulation of the microresonator refractive index can also be used to generate OFCs (see[15,18,19] and references therein). The FSR of a SBM is estimated as $\Delta\nu_{FSR} = c(2\pi n_0)^{-1}(r_0 R)^{-1/2}$ where $r_0$ is the fibre radius, $R \gg r_0$ is the axial radius, and $n_0$ is the refractive index of the microresonator material[23,24]. Remarkably, for SBMs, the value of $R$ can be dramatically large. For example, for the semi-parabolic SBM fabricated in ref.[28], the axial radius was $R \cong 0.7$ km and, potentially, can be an order of magnitude greater. A feasible FSR for a silica ($n_0 = 1.46$) SBM with $R = 10$ km and $r_0 = 100$ μm is $\Delta\nu_{FSR} \cong 33$ MHz. To maximise the frequency comb bandwidth, the SBM should have equally spaced axial eigenfrequencies. This feature is achievable with the SNAP technology as we can fabricate SBMs with parabolic or semi-parabolic shape[28]. Experimentally, OFCs generated by conventional bottle microresonators with axial radius $R$ of the order of 100 μm have been presented by several groups[29–31] starting with the first demonstration in ref.[32]. The generation of OFCs in parabolic SBMs by the nonlinear Kerr effect and by resonant harmonic parametric excitation has been theoretically studied[33,34]. However, to the best our knowledge, OFCs generated by parametric modulation of parabolic SBMs have not been experimentally demonstrated to date.

In this paper, we theoretically investigate the formation of OFCs by the parametric modulation of parabolic SBMs going beyond the lumped-element model[15,18–20,22] and taking into account the spatial distribution of the parametric modulation (SDPM). Based on the theory of SNAP microresonators with time-dependent parameters[35], we optimize the SDPM targeting to arrive at the maximum flat and broadband OFC spectrum. Remarkably, the characteristic ERV $\delta r_{par}$ of the SDPM can be much smaller than, as well as comparable to, the dramatically small nanometre-scale ERV $\Delta r_0$ of the SBM. Below, we demonstrate the critical importance of the SDPM in the adiabatic ($\nu_{par} \ll \Delta\nu_{FSR}$) and resonant ($\nu_{par} \approx 2\Delta\nu_{FSR}$) cases as well as for relatively small ($\delta r_{par} \ll \Delta r_0$) and large ($\delta r_{par} \sim \Delta r_0$) modulation amplitudes.

# Results

We consider a parabolic SBM formed by the nanoscale ERV of an optical fibre with radius $r_0$ illustrated in Fig. 1. Light is coupled into the SBM by a transverse optical waveguide placed at $z = z_0$. In our analysis, we model the input light as a monochromatic source $A_{in}(z,t) \cdot e^{-2\pi i \nu_{in} t}$ where function $A_{in}(z,t)$ is localised near point $z_0$ and has the characteristic switching time $\alpha^{-1}$:

$$A_{\text{in}}(z,t) = A_0 \delta(z - z_0) \begin{cases} (1 - e^{-\alpha t}) & t \geq 0 \\ 0 & t < 0 \end{cases}. \qquad (1)$$

If frequency $\nu_{in}$ is close to one of the cutoff frequencies of the optical fibre $\nu_c(z,t)$[23], the input light excites a WGM that slowly propagates along $z$ and bounces between the SBM turning points forming eigenmodes. The slowness of the propagating WGM depends on the proximity of its frequency $\nu_{in}$ to the cutoff frequency $\nu_c(z,t)$ which is assumed to have the parabolic shape perturbed by the parametric modulation:

$$\nu_c(z,t) = \nu_{c0} + \delta\nu_{par} \cdot \xi(z/L) \cdot \sin(2\pi\nu_{par} t) + \begin{cases} \Delta\nu_0 \left[ \left(\frac{z}{L}\right)^2 - 1 \right] & |z| \leq L \\ 0 & |z| > L \end{cases}. \qquad (2)$$

Here, $2L$ is the length of the SBM, $\nu_{c0}$ is the cutoff frequency of the uniform optical fibre away from the SBM, $\Delta\nu_0$ is the maximum of the cutoff frequency variation (CFV) that forms the SBM (Fig. 1), $\delta\nu_{par}$ is the maximum of the modulation amplitude, and $\xi(z/L)$ is its dimensionless spatial distribution.

For sufficiently small and slow deformation of the optical fibre, the WGM field distribution is separable in cylindrical coordinates as $e^{il\phi}\Theta_p(r)\Psi(z,t) \cdot e^{-i2\pi\nu_{in}t}$, where $l$, and $p$, are the azimuthal and radial quantum numbers. Then, under the condition that $\nu_c(z,t)$ is sufficiently close to $\nu_{in}$, function $\Psi(z,t)$ obeys the Schrödinger equation[35]:

$$i\frac{1}{2\pi\nu_{c0}}\partial_t\Psi = -\frac{1}{2\beta_{c0}^2}\partial_z^2\Psi + \frac{\nu_c(z,t) - \nu_{in} - i\gamma}{\nu_{c0}}\Psi - \frac{1}{\beta_{c0}^2}D_0\delta(z-z_0)\Psi + A_{in}(z,t). \qquad (3)$$

Here, $\beta_{c0} = 2\pi n_0 \nu_{c0}/c$ is the cutoff wavenumber, $n_0$ is the refractive index of the SBM, and $c$ is the speed of light. In Eq. (3), the absorption losses in the SBM are determined by the parameter $\gamma$ while the complex number $D_0$ takes into account the effect of the input-output waveguide coupled to the SBM at $z = z_0$[23].

In the presence of coupling and losses, the frequencies of the SBM are complex-valued and given by

$$\tilde{\nu}_q = \nu_q - i\Gamma_q = \left(\nu_{c0} - \Delta\nu_0 + \left(q + \frac{1}{2}\right)\Delta\nu_{\text{FSR}} - \delta\nu_{D_0,q}\right) - i\Gamma_q, \quad \Delta\nu_{\text{FSR}} = c/2\pi n_0 L \cdot \sqrt{2\Delta\nu_0/\nu_{c0}} \qquad (4)$$

where $q$ is the axial quantum number and $\Delta\nu_{\text{FSR}}$ is the axial FSR of the unmodulated SBM. Here $\delta\nu_{D_0,q}$ and $\Gamma_q$ are the frequency shift and width induced by coupling and material losses[23],

$$\delta\nu_{D_0,q} = \frac{c^2}{8\pi^2 n_0^2 \nu_{c0}} \text{Re}(D_0)|\varphi_q(z_0)|^2, \quad \Gamma_q = \gamma + \frac{c^2}{8\pi^2 n_0^2 \nu_{c0}} \text{Im}(D_0)|\varphi_q(z_0)|^2. \qquad (5)$$

In this equation, $\varphi_q(z)$ are the eigenmodes of Eq. (3) in the absence of source, parametric modulation, coupling effects, and losses ($A_0 = \delta v_{par} = D_0 = \gamma = 0$). From expression Eq. (5), the $Q$-factor of the eigenmode $q$ is calculated as

$$Q_q = v_{in}/\Gamma_q. \tag{6}$$

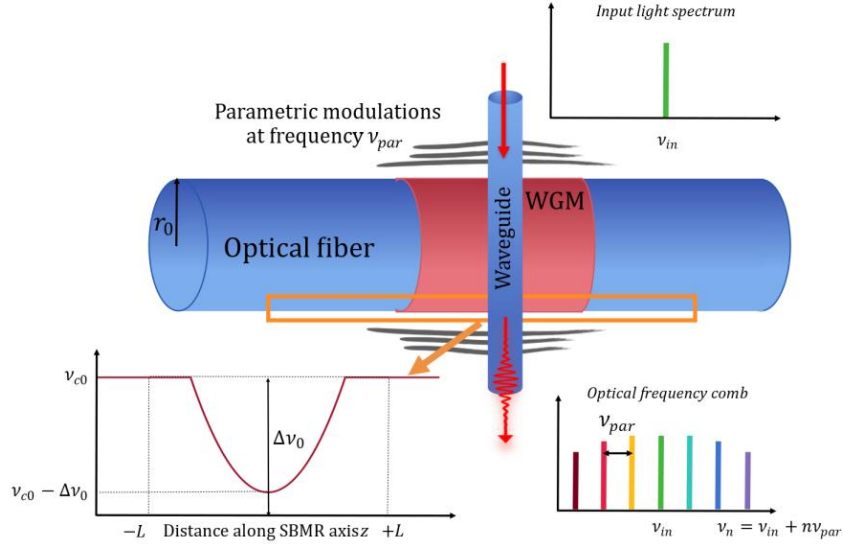

**Figure 1. Illustration of the considered system.** An SBM with parabolic CFV centred at $z_0 = 0$ introduced at the surface of an optical fibre with radius $r_0$. The SBM is coupled to the transverse waveguide that pumps light at frequency $v_{in}$ and excites WGMs. The SBM is parametrically modulated with frequency $v_{par}$ and generates an OFC at the output of the waveguide with the repetition rate $v_{par}$.

The OFC spectrum is determined from the solution of Eq. (3) as follows. First, we calculate the Fourier transform of this solution at the position $z = z_0$ of the source, $F(v, z_0, \delta v_{par}) = \int dt \exp(-2\pi i v t) \Psi(z_0, t)$. Here, for convenience, the modulation amplitude $\delta v_{par}$ from Eq. (2) is included. Next, we normalise $F(v, z_0, \delta v_{par})$ by relating it to the maximum of the output spectrum at zero modulation amplitude, $\max[|F(v, z_0, 0)|]$. Then, the output OFC power spectrum is found as

$$P(v) = \left|\frac{F(v,z_0,\delta v_{par})}{\max[|F(v,z_0,0)|]}\right|^2 \tag{7}$$

Below, we investigate the OFCs generated by the SBM determined by Eqs. (2) and (3) for the cases of adiabatically slow parametric modulation, $v_{par} \ll \Delta v_{FSR}$, and resonant parametric modulation, $v_{par} \approx 2\Delta v_{FSR}$.

### *Adiabatic parametric modulation of the SBM*

In this section, we consider the adiabatic modulation of the SBM assuming that the frequency $v_{par}$ is much smaller than the microresonator FSR, $v_{par} \ll \Delta v_{FSR}$, and the OFC resonances are well defined so that their width is much smaller than their separation, i.e., $\Gamma_q \ll v_{par}$. We set the input frequency $v_{in}$ equal to one of the SBM eigenfrequencies $v_{q_0}$. The simplest parametric modulation corresponds to an axially uniform SDPM with $\xi(z/L) = 1$ and $\varepsilon(z, t) = \delta\nu_{par} \cdot \sin(2\pi\nu_{par} t)$. In this case, the transmission power spectrum $P(\nu)$ vanishes at frequencies $\nu \neq \nu_{q_0} + n\nu_{par}$, $n = 0, \pm 1, \pm 2, ...$, and (see Methods)

$$P(\nu_{q_0} + n\nu_{par}) = \left| J_n\left(\frac{\delta\nu_{par}}{\nu_{par}}\right) J_0\left(\frac{\delta\nu_{par}}{\nu_{par}}\right) \right|^2. \tag{8}$$

From this equation, the OFC spectrum is formed by a series of equally spaced resonances with magnitudes determined by the dependence of Bessel functions $J_n\left(\frac{\delta\nu_{par}}{\nu_{par}}\right)$ on the comb resonance number $n$. From the asymptotic of the Bessel function for large order and argument, $|n|, \frac{\delta\nu_{par}}{\nu_{par}} \gg 1$, we find that the OFC spectrum determined by Eq. (8) is relatively flat within the bandwidth including $N \sim \frac{2\delta\nu_{par}}{\nu_{par}}$ resonances.

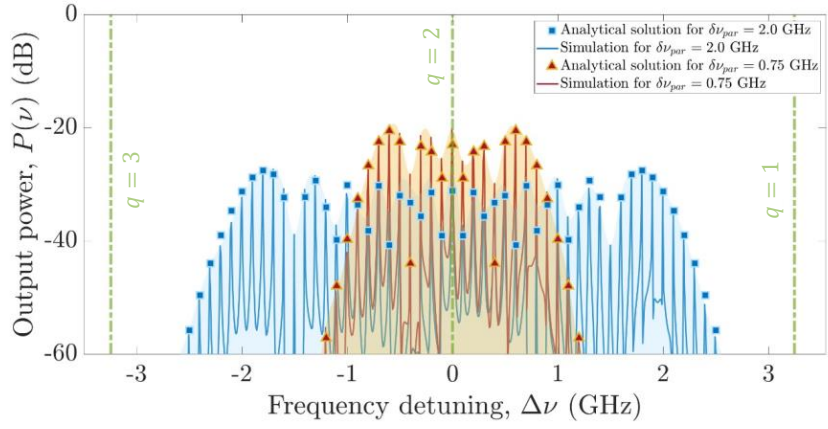

**Figure 2. OFCs generated by the spatially uniform adiabatic modulation of a parabolic SBM.** The blue and red lines correspond to the OFC spectrum obtained from numerical solution of Eq. (3) for constant modulation amplitudes $\delta\nu_{par} = 2$ GHz and $\delta\nu_{par} = 0.75$ GHz, respectively. In both cases, the Q-factor is $Q_{q_0} = 1.2 \times 10^7$. The blue squares and the red triangles correspond to the spectrum found from Eq. (8) for $\delta\nu_{par} = 2$ GHz and $\delta\nu_{par} = 0.75$ GHz, respectively.

The comparison of the OFC spectrum obtained from the numerical solution of Eq. (3) with that given by Eq. (8) and shown in Fig. 2 demonstrates their excellent agreement. We consider an SBM with maximum CFV $\Delta\nu_0 = 20$ GHz, total length $2L = 284$ μm and radius $r_0 = 20$ μm. The input-output waveguide is placed at the centre of the SBM, $z_0 = 0$. The input light frequency is set to $\nu_{in} = 200$ THz which is equal to the SBM eigenfrequency $\nu_{q_0}$ with axial quantum number $q_0 = 2$. For these parameters, the axial FSR of the SBM determined by Eq. (4) is $\Delta\nu_{FSR} = 3.25$ GHz. We analyse the OFCs generated with two different relatively small modulation amplitudes, $\delta\nu_{par} = 0.1\Delta\nu_0 = 2$ GHz and $\delta\nu_{par} = 0.04\Delta\nu_0 = 0.75$ GHz, at frequency $\nu_{par} = 100$ MHz. From the scaling relation $\delta r_{par} = r_0 \delta\nu_{par}/\nu_{c0}$, these amplitudes correspond to the ERV $\delta r_{par} = 200$ pm and $\delta r_{par} = 75$ pm, respectively, while the total ERV of the SBM considered is $\Delta r_0 = r_0 \Delta\nu_0/\nu_{c0} = 2$ nm. The material attenuation and coupling coefficient are set to $\gamma = 2\pi$ MHz and $D_0 = 0.001(1+i)$ $\mu m^{-1}$, which corresponds to $Q_{q_0} = 1.2 \times 10^7$ calculated from Eqs. (5) and (6).

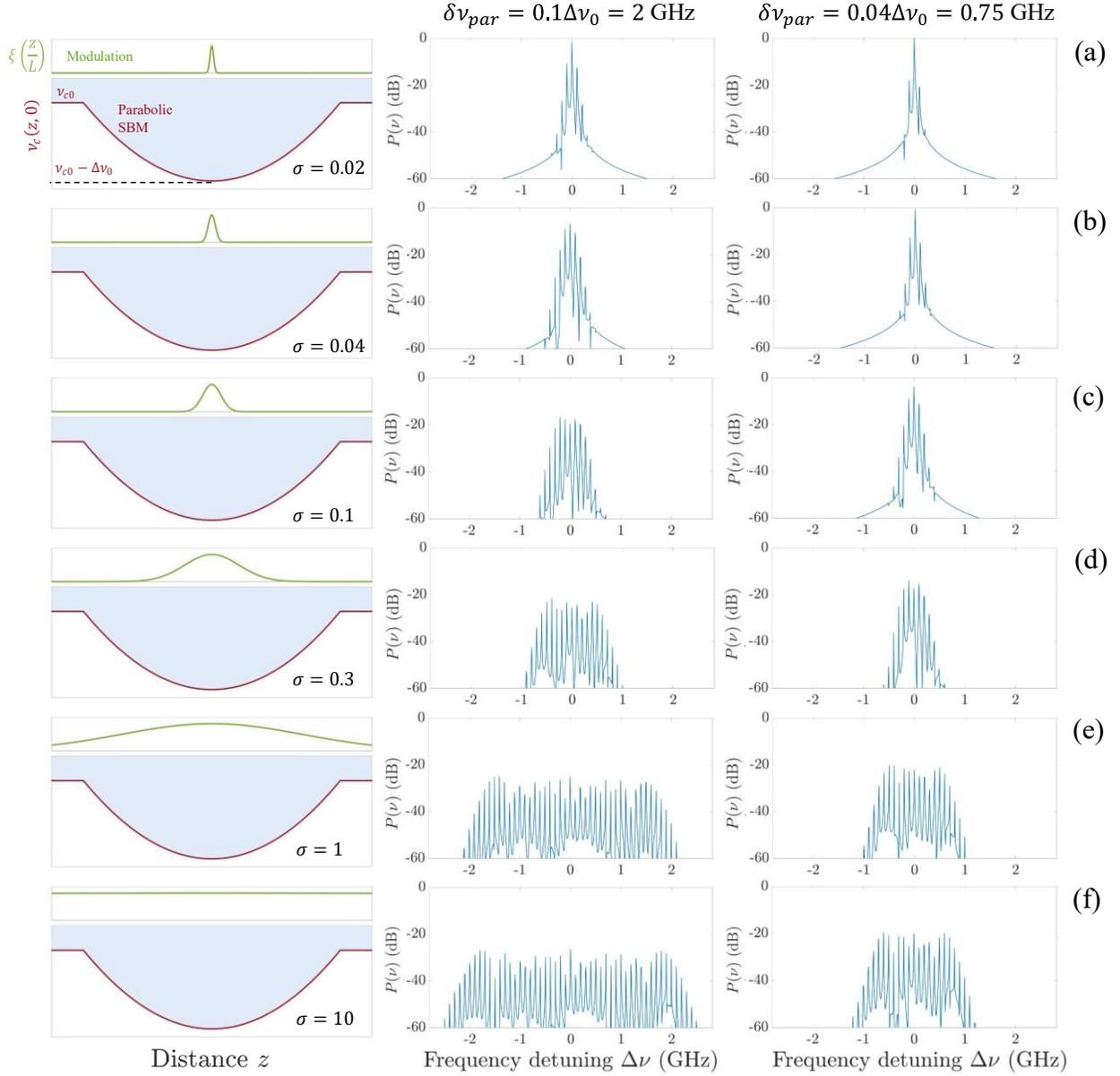

**Figure 3. OFCs generated by the adiabatic modulation of parabolic SBM with Gaussian SDPMs.** The SDPM is defined by $\xi(z/L) = e^{-(z/\sigma \cdot L)^2}$, where $\sigma$ is the dimensionless SDPM width and $L$ is the SBM length. Left column shows CFV and $\xi(z/L)$ for different values of $\sigma$ varying from (a) $\sigma = 0.02$ to (f) $\sigma = 10$. Two values of the modulation amplitude maximum are considered: $\delta v_{par} = 0.1 \Delta v_0 = 2$ GHz (centre column) and $\delta v_{par} = 0.04 \Delta v_0 = 0.75$ GHz (right column). For $\delta v_{par} = 0.75$ GHz, the OFC has a smaller bandwidth, though the power level is $\sim 10$ dB higher than in the case of $\delta v_{par} = 2$ GHz. In both cases, a limited number of comb resonances are formed at $\sigma = 0.02$, when the SDPM is strongly localized near the centre of the SBM. In (a)-(f), the OFC spectrum becomes wider and flatter with growing $\sigma$ and achieves its optimal shape for a uniform SDPM in (f).

We now investigate the effect of the SDPM, $\xi(z/L)$, on the generation of OFCs. We assume the Gaussian SDPM, $\xi(z/L) = e^{-(z/\sigma L)^2}$, where the dimensionless parameter $\sigma$ determines the ratio of characteristic SDPM width and SBM length $L$. Excitation of a broader OFC spectrum at the minimum required power depends on the techniques to generate the parametric modulations. In particular, the required input power may be determined by the maximum amplitude of modulation, its integrated intensity, or by a more complex functional dependence of the modulation parameters on the input power. Here, we suggest that the required power is determined by the maximum amplitude of the SDPM and set the same $\delta\nu_{par}$ for all the cases considered.

Fig. 3 shows the results of our numerical simulations for SBM with the parameters indicated above and modulation amplitude maxima $\delta\nu_{par} = 2$ GHz and $\delta\nu_{par} = 0.75$ GHz, which are much smaller that the full CFV of the SBM $\Delta\nu_0 = 20$ GHz. The graphs in the first column of this figure illustrate the parabolic CFV (red solid line) and the Gaussian SDPM (green solid line) for different values of $\sigma$. The plots in the second and third columns of Fig. 3 show the OFC spectrum of the signal obtained from Eq. (3) for the SDPM illustrated in the first column for the modulation amplitude maxima $\delta\nu_{par} = 2$ GHz and $\delta\nu_{par} = 0.75$ GHz, respectively. In the case $\sigma = 0.02$, the Gaussian SDPM is localized near the centre of the SBM as shown in Fig. 3(a) and the OFC spectrum exhibits only a few comb resonances. The spectrum of the frequency combs becomes wider and flatter with growing $\sigma$ (Fig. 3(a)-(f)). In the case $\sigma = 10$ shown in Fig. 3(f), the parametric modulation has close to the uniform SDPM over the whole length of the SBM and the spectrum coincides with that obtained from Eq. (8) and illustrated in Fig. 2.

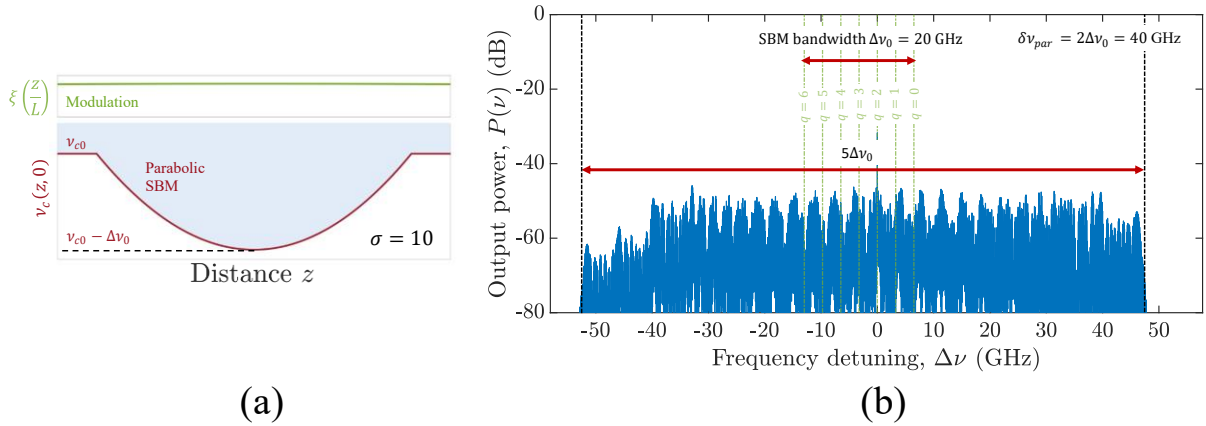

(a)           (b)

**Figure 4. OFC generated by the adiabatic modulation of a parabolic SBM with uniform SDPM having a large modulation amplitude $\delta\nu_{par} = 2\Delta\nu_0 = 40$ GHz.** (a) Illustration of the uniform SDPM and parabolic CFV. (b) The generated OFC.

So far, we considered the modulation of the SBM with maximum amplitudes $\delta\nu_{par}$ much smaller than the full CFV $\Delta\nu_0 = 20$ GHz of the resonator. However, the dramatically small CFV and corresponding ERV of the SBM ($\Delta r_0 = 2$ nm in the cases considered) makes the modulations comparable or larger than the resonator CFV realistic as well. As an example, Fig. 4 shows the plot of the OFC generated by the uniform SDPM with $\delta\nu_{par} = 2\Delta\nu_0 = 40$ GHz. The corresponding ERV modulation amplitude is 4 nm (compare it with the 8 nm ERV oscillation amplitude experimentally achieved in ref.[22] and discussed in the introduction). It is seen from Fig. 4(b) that, as expected, the OFC spectrum expands over the bandwidth $5\Delta\nu_0$ (from $\nu_{q_0} - 2.5\Delta\nu_0$ to $\nu_{q_0} + 2.5\Delta\nu_0$). The number of OFC resonances within the central most uniform part of this spectrum, from -40 to 40 GHz, is well estimated by $N \sim \frac{2\delta\nu_{par}}{\nu_{par}} = \frac{4\Delta\nu_0}{\nu_{par}} = 800$, followed from the asymptotic of Eq. (8) discussed above, while the full bandwidth includes close to $\frac{5\Delta\nu_0}{\nu_{par}} = 1000$ OFC resonances.

### *Resonant parametric modulation of the SBM*

As in the previous section, we place the input light source in the middle of the SBM ($z_0 = 0$) so that the excited eigenmodes have even parity. Then, due to the symmetry of the system, transitions between eigenmodes of different parity are suppressed and the parametric modulation frequency has to be $\nu_{par} \approx 2\Delta\nu_{FSR}$ to achieve an effective resonant excitation of the SBM eigenmodes. In this case, the SBM has the maximum CFV $\Delta\nu_0 = 5$ GHz, the total length $2L = 9.18$ mm, and radius $r_0 = 80$ $\mu$m. The axial FSR of this SBM is $\Delta\nu_{FSR} = 50$ MHz, so the resonant condition takes place at $\nu_{par} \approx 2\Delta\nu_{FSR} = 100$ MHz. The input light frequency is set to $\nu_{in} = \nu_{q_0} = 200$ THz, where the axial mode number is $q_0 = 50$. The attenuation factor and coupling coefficient are set to $\gamma = 2\pi$ MHz and $D_0 = 0.001(1+i)$ $\mu m^{-1}$. The corresponding Q-factor of mode $q_0$ calculated from Eqs. (5) and (6) is $Q_{q_0} = 3 \times 10^7$.

The results of our numerical modelling are shown in Fig. 5. As in the adiabatic case, the OFCs are generated by the Gaussian SDPM determined by $\xi(z/L) = e^{-(\sigma \cdot z/L)^2}$ with dimensionless widths $\sigma$ varying from 0.02 to 10. However, contrary to the adiabatic case, the OFC spectral bandwidth behaves nonmonotonically as a function of $\sigma$. The corresponding SDPM and CFV plots are shown in the left-hand column of Fig. 5. The centre and right-hand columns of this figure show the OFC power spectra for two modulation amplitudes, $\delta\nu_{par} = 200$ MHz and $\delta\nu_{par} = 75$ MHz. From the scaling relation $\delta r_{par} = r_0 \delta\nu_{par}/\nu_{c0}$, these amplitudes correspond to the ERV $\delta r_{par} = 80$ pm and $\delta r_{par} = 30$ pm, respectively. From the similar scaling relation, the ERV of the SBR considered is $\Delta r_0 = 2$ nm, as in the adiabatic case. It is seen from Fig. 5 that, with increasing the SDPM width from very narrow at $\sigma = 0.02$ (Fig. 5(a)), the OFC spectral bandwidth first becomes wider (Figs. 5(b)-(c)), reaches the optimum profile at around $\sigma = 0.3$ (Fig. 5(d)), and then shrinks down (Fig. 5(e)) and vanishes for the practically uniform SDPM at $\sigma = 10$ (Fig. 5(f)). Vanishing of the OFC spectral bandwidth for narrow SDPM (Fig. 5(a)) is straightforwardly explained by the reduction of integrated modulation power. The less obvious suppression of the OFC generation for the SDPM approaching the uniform distribution (Fig. 5(f)) can be clarified with the help of the quantum and semiclassical perturbation theory. Indeed, the transition between SBM axial modes with different quantum numbers $q$ caused by the spatially uniform perturbation is strongly suppressed due to the orthogonality of these modes.

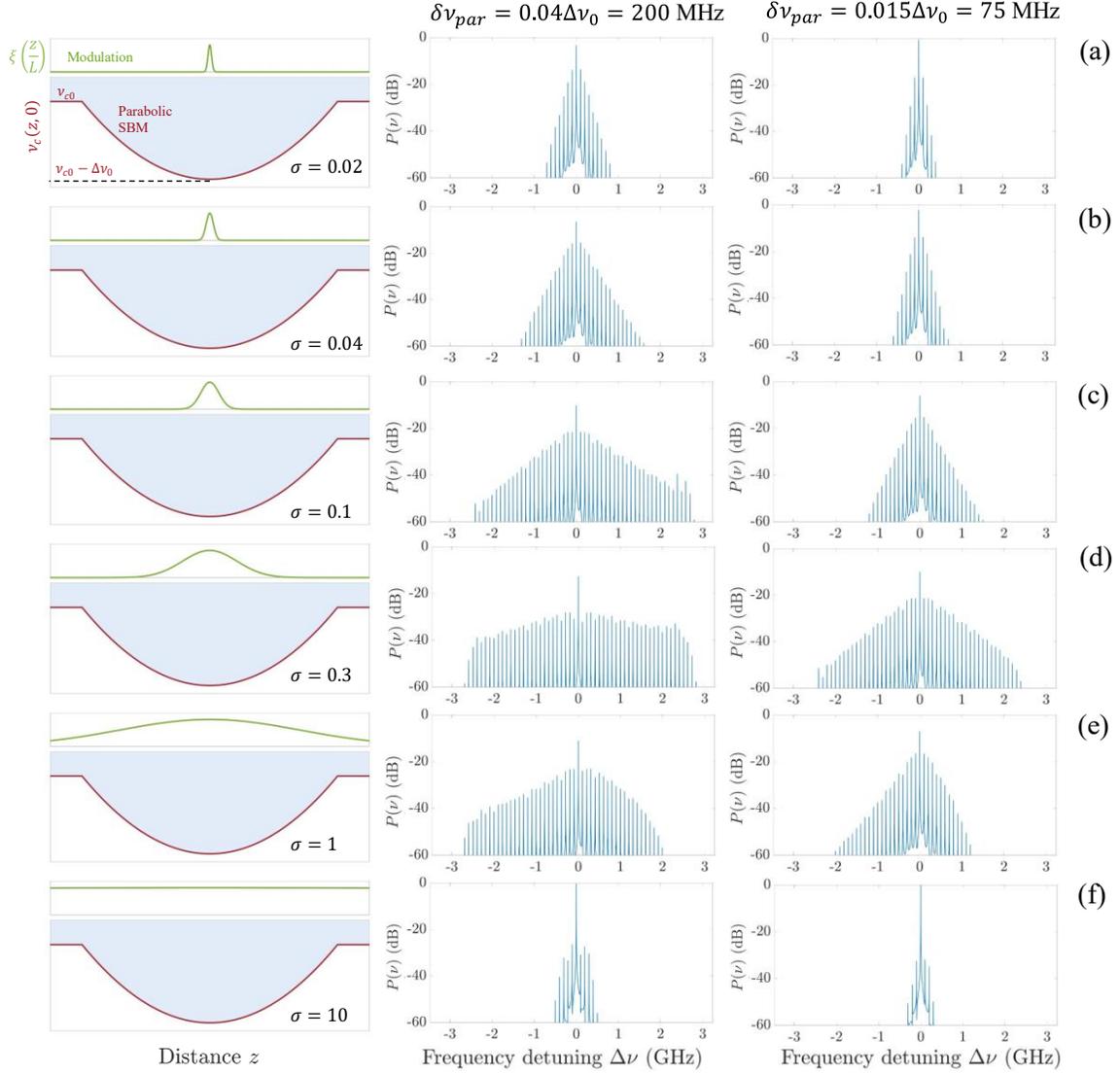

**Figure 5. Frequency combs generated by resonant parametric modulations with Gaussian SDPMs.** Two maximum modulation amplitudes, $\delta\nu_{par} = 200$ MHz (centre column) and $\delta\nu_{par} = 75$ MHz (right column) are considered for the parabolic CFV and SDPMs with $\xi(z/L) = e^{-(\sigma \cdot z/L)^2}$ for different dimensionless widths $\sigma$ (left column). (a) Strongly localized modulation with $\sigma = 0.02$ leads to the formation of a narrow bandwidth OFC spectrum with a few comb resonances. (b)-(c) The bandwidth of the OFC spectrum increases with the SDPM width $\sigma$. (d) The optimal OFC spectrum with largest bandwidth is achieved at $\sigma \sim 0.3$. (f) Generation of combs is strongly suppressed for the uniform SDPM.

Similar to the previous section, in addition to the cases of relatively small modulation amplitudes $\delta\nu_{par} \ll \Delta\nu_0$, it is interesting to investigate the case when $\delta\nu_{par}$ is comparable to the full CFV $\Delta\nu_0$. As an example, Fig. 6 compares the OFC spectra generated by the same SBM modulated with maximum amplitude $\delta\nu_{par} = 2\Delta\nu_0 = 10$ GHz and Gaussian profile $\xi(z/L) = e^{-(z/\sigma \cdot L)^2}$. Fig. 6(a) shows the case with $\sigma = 0.3$ (optimal for the relatively weak modulation, see Fig. 5(d)) and Fig. 6(b) shows the case of close to uniform SDPM with $\sigma =$

10. It is seen that, contrary to the case of relatively small modulation amplitude, the uniform modulation generates an OFC which is, in general, comparable with that for $\sigma = 0.3$. However, the OFC spectrum for $\sigma = 0.3$ is closer to uniform and on average has a greater power over the larger bandwidth. We suggest that an OFC profile with better flatness and larger bandwidth can be achieved with an optimization using a more complex SDPM. The detailed solution of this problem is beyond the scope of this paper.

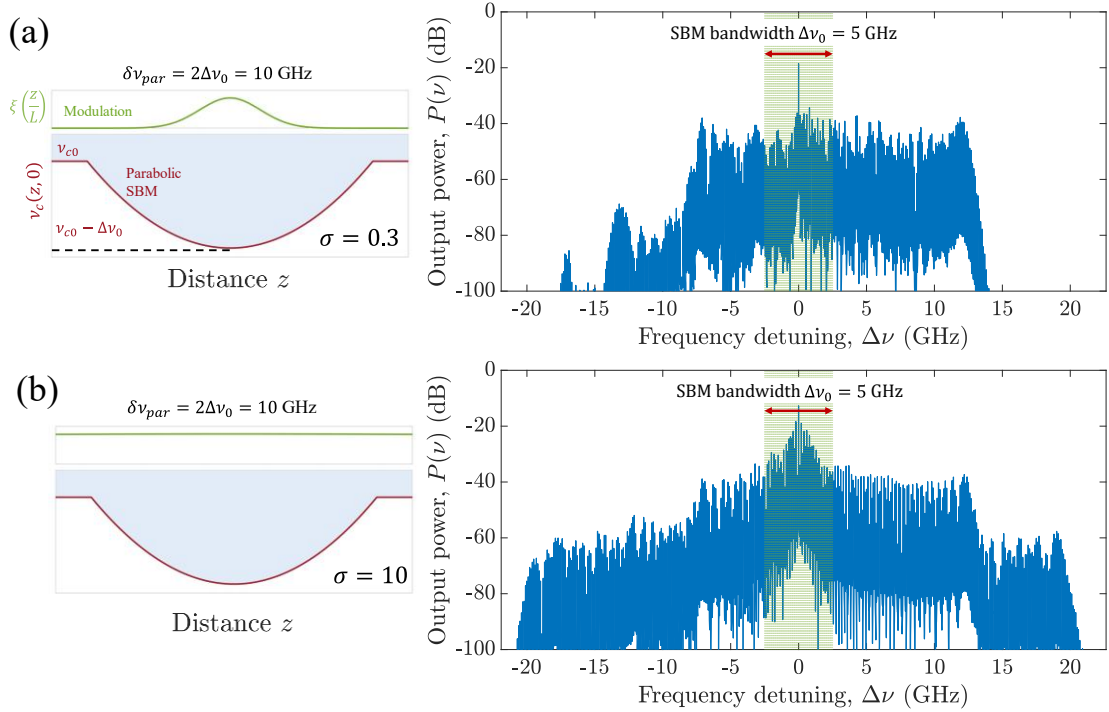

**Figure 6. OFCs generated by resonant parametric modulations with Gaussian SDPMs and large modulation amplitude $\delta\nu_{par} = 2\Delta\nu_0 =$10 GHz.** Parabolic CFV and Gaussian modulation profiles $\xi(z/L) = e^{-(z/\sigma \cdot L)^2}$ (left column) and corresponding OFC (right column) for (a) $\sigma = 0.3$ and (b) $\sigma = 10$.

**Conclusion**

Parametric modulation of optical microresonators is a promising approach for the generation of OFCs which has been demonstrated in spherical[15], ring (racetrack)[19,20], and toroidal[15,18] microresonators. The repetition rate of the OFC excited by these resonators is inverse proportional to their perimeter so that the OFCs with the characteristic repetition rate smaller than a gigahertz correspond to resonators with macroscopically large dimensions (see the Introduction for details). However, it was proposed in refs.[33,34,36] that an SBM with much smaller dimensions can be used to generate OFCs with repetition rate below a gigahertz.

In this paper, we consider the generation of OFCs in a parabolic SBM by the parametric modulation of its parameters. The main difference of an SBM compared to spherical and toroidal microresonators consists in its strong axial elongation. For this reason, the spatial distribution of the parametric modulation (SDPM) along the SBM axis becomes of major importance. Furthermore, the characteristic parametric ERV modulation $\delta r_{par}$ can be comparable or larger than the dramatically small nanometre-scale ERV $\Delta r_0$ of the SBMs considered. In our numerical simulations, we investigate the effect of the SDPM assuming that it has the Gaussian shape. In the adiabatic case, when the modulation frequency is much smaller than the axial FSR, $\nu_{par} \ll \Delta\nu_{FSR}$, we show that a uniform SDPM produces the OFC with the best flatness and largest spectral bandwidth (Fig. 3(f)). An adiabatic SDPM with uniform amplitude $\delta r_{par} \geq \Delta r_0$ generates a relatively flat OFC with bandwidth proportional to this amplitude (Fig. 4). In the case of resonant modulation, $\nu_{par} \approx 2\Delta\nu_{FSR}$, and relatively small modulation amplitude, $\delta r_{par} \ll \Delta r_0$, the generated OFC vanishes both for very narrow (Fig. 5(a)) and uniform (Fig. 5(f)) SDPM and achieves the optimal shape with the best flatness and largest bandwidth at the SDPM with intermediate width (Fig. 5(d)). For larger modulation amplitudes, $\delta r_{par} \geq \Delta r_0$, the resonant modulation with a nonuniform SDPM results in a flater OFC with a broader bandwidth compared to that generated by the uniform SDPM (Fig. 6). We suggest that an SDPM with a more complex dependence on the axial coordinate, which takes into account the actual axial distribution of the SBM modes, may lead to a better OFC optimization.

The optimized generation of OFCs by an SBM proposed in this paper can be realized experimentally by one of the following approaches. First, modulation of the refractive index and radiation pressure of an SBM can be induced by the input resonant pump light harmonically modulated with frequency $\nu_{par}$. To ensure the resonant enhancement of the pump light, its spectrum should be localized near an SBM cutoff frequency $\nu_{cp}$. While the case when the pump light induces the OFC directly (i.e., when $\nu_{cp} = \nu_{c0}$, see Eq. (2)) is of special interest, we assumed in this paper that the cutoff frequencies $\nu_{cp}$ and $\nu_{c0}$ (as well as the spectrum of the pump light and the frequency $\nu_{in}$ of a relatively weak input light) are separated so that the spectrum of pump light and the spectrum of the generated OFC do not overlap. The required spatial distribution of the pump light can be achieved by adjusting its spectrum. Second, the adiabatic excitation of the OFCs can be performed in a way similar to that demonstrated in ref.[22] by excitation of acoustic modes of an SBM with a strong pump light. Third, assuming that the ultraprecise fabrication of an SBM of material with strong second-order nonlinearities may become possible, the refractive index modulation can be performed by applying a periodic electric field to the SBM fabricated of such material[18]. Finally, an SBM can be fabricated at the surface of a microcapillary filled with such highly nonlinear or piezoelectric material, which can be used to mechanically modulate the SBM parameters[37].

## Acknowledgements

The research of M.C. and M.S. was supported by the EPSRC, grants EP/P006183/1 and EP/W002868/1. The research performed by A.M. was carried out at the JPL, Caltech, under a contract with the NASA (80NM0018D0004).

## Methods

### Generation of OFCs by a spatially uniform parametric modulation

We consider a cutoff frequency that depends on time and position as $v_c(z,t) = v_{c0} + \Delta v(z) + \delta v_{par} \cdot \sin(2\pi v_{par} t)$, where $|\Delta v(z)| \ll v_{c0}$ and $\delta v_{par} = const$. After the substitution

$$\Psi(z,t) = \psi(z,t) e^{\frac{i\delta v_{par}}{v_{par}} \cos(2\pi v_{par} t)}, \tag{9}$$

Eq. (3) can be written as

$$\left[ i \frac{1}{2\pi} \partial_t - \hat{L} \right] \psi(z,t) = S(z,t) \tag{10a}$$

$$\hat{L} = -\frac{1}{2}\kappa \partial_z^2 + (v_{c0} - v_{in} + \Delta v(z) - i\gamma) - \kappa D_0 \delta(z - z_0) \tag{10b}$$

$$S(z,t) = v_{c0} A_{in}(z,t) e^{-i\frac{\delta v_{par}}{v_{par}} \cos(2\pi v_{par} t)} \tag{10c}$$

where $\kappa = v_{c0}/\beta_{c0}^2$. For the parabolic CFV, $\Delta v(z) = \Delta v_0 \left[ \left(\frac{z}{L}\right)^2 - 1 \right]$, we search for solution of Eq. (10a) in the form

$$\psi(z,t;z_0) = \sum_q \varphi_q(z) \varphi_q^*(z_0) a_q(t) \tag{11}$$

where $a_q(t)$ are functions of time to be determined and $\varphi_q(z)$ are the normalised eigenfunctions of the operator $\hat{L}$ with associated eigenfrequencies $\varsigma_q$. For finite material loss $\gamma$ and coupling $D_0$ considered, the eigenfrequencies are complex-valued and equal to $\varsigma_q = v_q - v_{in} - i\Gamma_q$ (see Eqs. (4) and (5)). Nevertheless, due to the relatively small values of $\gamma$ and $\text{Im}(D_0)$ assumed, the normalization of $\varphi_q(z)$ and their orthogonality can be achieved with the required accuracy.

In order to determine $a_q(t)$, we substitute Eq. (11) into Eq. (10a),

$$\sum_q \varphi_q(z) \varphi_q^*(z_0) \left[ i \frac{1}{2\pi} \frac{d}{dt} a_q(t) - \varsigma_q a_q(t) \right] = S(z,t), \tag{12}$$

multiply each side of Eq. (12) by $\varphi_{q'}^*(z)$ and integrate over $z$. As a result, we arrive at the following equations for $a_q(t)$:

$$\frac{1}{2\pi} \frac{d}{dt} a_q(t) + i\varsigma_q a_q(t) = -iv_{c0} A_0 (1 - e^{-\alpha t}) e^{-i\frac{\delta v_{par}}{v_{par}} \cos(2\pi v_{par} t)} \tag{13}$$

with initial conditions $a_q(t=0) = A_{in}(z,0) = 0$. Solving these equations and application of the Jacobi-Anger expansion[38], results in the following solution of Eq. (3) at $z = z_0$:

$$\Psi(z_0,t) = v_{c0} A_0 \sum_q |\varphi_q(z_0)|^2 \sum_n \sum_m (-1)^m i^n J_{n-m}\left(\frac{\delta v_{par}}{v_{par}}\right) J_m\left(\frac{\delta v_{par}}{v_{par}}\right) \frac{e^{2\pi i n v_{par} t}}{(v_q - v_{in} + m v_{par}) - i\Gamma_q}. \tag{14}$$

We assume now that the input frequency $v_{in}$ is situated in a small vicinity of one of the SBM frequencies $v_{q_0}$ of the order of the resonance width, $|v_{in} - v_{q_0}| \sim \Gamma_{q_0}$, and the excited OFC resonances are well defined so that $\Gamma_{q_0} \ll \nu_{par}$. In addition, we assume that none of the SBM frequencies $\nu_q$, except for $\nu_{q_0}$, are close to the parametrically modulated OFC resonances. Such situation may occur for adiabatic modulation, when $\nu_{par} \ll \Delta\nu_{FSR}$. Then, Eq. (14) is reduced to

$$\Psi(z_0, t) = \sum_n \Psi_n e^{2\pi i n \nu_{par} t}, \tag{15}$$

where

$$\Psi_n = i^{n+1} \frac{\nu_{c0}}{\Gamma_{q_0}} A_0 |\varphi_{q_0}(z_0)|^2 J_n\left(\frac{\delta\nu_{par}}{\nu_{par}}\right) J_0\left(\frac{\delta\nu_{par}}{\nu_{par}}\right). \tag{16}$$

In particular, at zero modulation, we have:

$$\Psi(z_0, t) = \Psi_0 = i \frac{\nu_{c0}}{\Gamma_{q_0}} A_0 |\varphi_{q_0}(z_0)|^2. \tag{17}$$

Substitution of Eqs. (16), (17) into Eq. (7) yields Eq. (8).

## Numerical solution of the Schrödinger equation

The Schrödinger equation is solved using the Fourier split-step method[39], which can be briefly described as follows. We first rewrite Eq. (3) as[40]:

$$i\partial_\tau \widehat{\Psi} = -\frac{1}{2}\partial_x^2 \widehat{\Psi} + \frac{1}{2}V_0\left[x^2 + \delta\hat{\nu}_{par} x_c^2 \xi(x/x_c)\right]\widehat{\Psi} + \Delta\hat{\nu}\widehat{\Psi} - i\hat{\gamma}\widehat{\Psi} - \widehat{D}_0\delta(x - x_0)\widehat{\Psi} + \hat{A}_{in}(x, \tau) \tag{18}$$

In this equation, we have introduced the dimensionless variables $\tau = t/T_0$, $x = z/L_0$ and $\widehat{\Psi} = \Psi/I_0$, where $T_0$, $L_0$ and $I_0$ are the scaling factors. The dimensionless parameters in Eq. (18) are defined as $V_0 = 8\pi^2 \Delta\nu_0 \nu_{c0} n_0^2 L_0^4/c^2 L^2$, $\delta\hat{\nu}_{par} = \delta\nu_{par}/\Delta\nu_0$, $x_c = L/L_0$, $\Delta\hat{\nu} = 2\pi(\Delta\nu_0 + \nu_{c0} - \nu_{in})T_0$, $\hat{\gamma} = 2\pi\gamma T_0$, $\widehat{D}_0 = L_0 D_0$ and $\hat{A}_{in}(x, \tau) = 2\pi\nu_{c0}T_0/L_0 I_0 \cdot A_{in}(x, \tau)$.

We can split the r.h.s. of Eq. (18) into the momentum space component

$$\widehat{D} = -\frac{1}{2}\partial_x^2 \tag{19}$$

and the position space component

$$\widehat{N} = \frac{1}{2}V_0\left[x^2 + \delta\hat{\nu}_{par} x_c^2 \xi(x/x_c)\right] + \Delta\hat{\nu} - i\hat{\gamma} - \widehat{D}_0\delta(x - x_0) \tag{20}$$

and rewrite Eq. (18) in terms of these two operators as:

$$i\partial_\tau \widehat{\Psi} = [\widehat{D} + \widehat{N}]\widehat{\Psi} + \hat{A}_{in}(x, \tau) \tag{21}$$

For a small time step $\Delta\tau$, the solution of this equation can be approximated as

$$\widehat{\Psi}(\tau + \Delta\tau) = e^{-i\Delta\tau \widehat{D}} e^{-i\Delta\tau \widehat{N}} \widehat{\Psi}(\tau) - i\Delta\tau \hat{A}_{in}(x, \tau) \tag{22}$$

with an error of the order of $\Delta\tau^2$. To compute the term associated to the operator $\widehat{D}$, we use the spatial fast Fourier transform (FFT) of the field $\widehat{\Psi}$:

$$\widehat{D}\widehat{\Psi} = FFT^{-1}\left\{e^{-\frac{i}{2}k^2} FFT(\widehat{\Psi})\right\} \tag{23}$$

where $k$ represents the coordinate in the reciprocal space. We solve Eq. (22) with a uniform space grid of $N = 2^n$ points defined as $x_n = n\Delta x$, with $n \in [-N/2, N/2 - 1]$. The points in the reciprocal space are defined as $k_n = 2\pi n/\Delta x N$.